\documentclass[10pt,twocolumn,english,showpacs,preprintnumbers,amsmath,amssymb,floatfix,nofootinbib]{revtex4-1} 

\usepackage[normalem]{ulem}   
\usepackage{caption}
\usepackage{tikz,xcolor}
\usepackage[colorlinks = true,
linkcolor = blue,
urlcolor  = blue,
citecolor = blue,
anchorcolor = blue]{hyperref}

\definecolor{lime}{HTML}{A6CE39}
\DeclareRobustCommand{\orcidicon}{%
	\begin{tikzpicture}
		\draw[lime, fill=lime] (0,0)
		circle [radius=0.16]
		node[white] {{\fontfamily{qag}\selectfont \tiny ID}};
		\draw[white, fill=white] (-0.0625,0.095)
		circle [radius=0.007];
	\end{tikzpicture}
	\hspace{-2mm}
}
\foreach \x in {A, ..., Z}{%
\expandafter\xdef\csname orcid\x\endcsname{\noexpand\href{https://orcid.org/\csname orcidauthor\x\endcsname}{\noexpand\orcidicon}}
}

\usepackage[T1]{fontenc}
\usepackage[latin9]{inputenc}
\usepackage{color}
\usepackage{array}
\usepackage{amstext}
\usepackage{subfig}
\usepackage{graphicx}
\usepackage{esint}
\usepackage{rotating}
\usepackage{slashed}
\usepackage{siunitx}
\usepackage[normalem]{ulem}
\usepackage[font=small,labelfont=bf]{caption}
\usepackage{lipsum}
\usepackage{xcolor}
\usepackage{comment}
\usepackage{amssymb}
\usepackage{booktabs} 

\newcommand{\xpom}{x_{\xpom}}

\makeatletter

\@ifundefined{textcolor}{}
{%
 \definecolor{BLACK}{gray}{0}
 \definecolor{WHITE}{gray}{1}
 \definecolor{RED}{rgb}{1,0,0}
 \definecolor{GREEN}{rgb}{0,1,0}
 \definecolor{BLUE}{rgb}{0,0,1}
 \definecolor{CYAN}{cmyk}{1,0,0,0}
\definecolor{MAGENTA}{cmyk}{0,1,0,0}
 \definecolor{YELLOW}{cmyk}{0,0,1,0}
 }

\@ifundefined{definecolor}
{\usepackage{color}}{}
\@ifundefined{definecolor}
{\usepackage{color}}{}
\makeatother
\usepackage{babel}

\def\Re{{\cal R \mskip-4mu \lower.1ex \hbox{\it e}\,}}
\def\Im{{\cal I \mskip-5mu \lower.1ex \hbox{\it m}\,}}

\def\tev{\,{\ifmmode\mathrm {TeV}\else TeV\fi}}
\def\gev{\,{\ifmmode\mathrm {GeV}\else GeV\fi}}
\def\mev{\,{\ifmmode\mathrm {MeV}\else MeV\fi}}
\def\to{\rightarrow}

\begin{document}

%
%
\title{Improved Constraints on Pion Fragmentation Functions from Simulated Electron-Ion Collider Data} 
%
%
\author{Maryam~Soleymaninia$^{1,2}$\orcidB{}}
\email{Maryam\_Soleymaninia@ipm.ir}

\author{Hamzeh~Khanpour$^{3,4,1}$\orcidE{}}
\email{Hamzeh.Khanpour@cern.ch}

\author{Majid~Azizi$^{1}$\orcidD{}}
\email{Ma.Azizi@ipm.ir} 

\author{Hadi~Hashamipour$^{5}$\orcidA{}}
\email{Hadi.Hashamipour@lnf.infn.it}

\affiliation {
$^{1}$School of Particles and Accelerators, Institute for Research in Fundamental Sciences (IPM), P.O.Box 19395-5531, Tehran, Iran.   \\
$^{2}$Department of Computer Engineering, UAE.C., Islamic Azad University, Dubai, United Arab Emirates\\
$^{3}$AGH University, Faculty of Physics and Applied Computer Science, Al. Mickiewicza 30, 30-055 Krakow, Poland. \\ 
$^{4}$Department of Physics, University of Science and Technology of Mazandaran, P.O.Box 48518-78195, Behshahr, Iran.  \\
$^{5}$Istituto Nazionale di Fisica Nucleare, Gruppo collegato di Cosenza, I-87036 Arcavacata di Rende, Cosenza, Italy.   
}

\date{\today}

%
\begin{abstract}

We present a quantitative assessment of the anticipated impact of future Electron-Ion Collider (EIC) 
measurements on the extraction of parton-to-pion fragmentation functions (FFs). Our analysis 
combines simulated semi-inclusive deep-inelastic scattering (SIDIS) pseudo-data at EIC energies of 45 GeV and 140 GeV 
with existing single-inclusive electron-positron annihilation (SIA) and SIDIS experimental data. 
The pion fragmentation functions are determined at next-to-leading-order (NLO) accuracy using a perturbative QCD 
framework and a neural network parametrization. Uncertainties are rigorously estimated through Monte Carlo 
sampling, accounting for both experimental errors and variations in input parton distribution functions. 
Our results demonstrate that incorporating EIC pseudo-data reduces their uncertainties, especially at medium to large momentum fractions ($z$).
This improvement is particularly pronounced for the gluon and selected quark FFs, highlighting the 
substantial role that EIC measurements will play in achieving high-precision extractions of FFs and 
informing future experimental and theoretical developments in collider physics.

\end{abstract}
%

\maketitle
\tableofcontents{}

%
\section{Introduction} \label{sec:introduction} 
%

Fragmentation functions (FFs) are fundamental components of quantum chromodynamics (QCD), describing the non-perturbative 
transition of partons into observed hadrons. These functions serve as critical inputs for theoretical predictions 
across various high-energy collision processes, including 
single-inclusive electron-positron annihilation (SIA), semi-inclusive deep-inelastic scattering (SIDIS), and 
hadron-hadron collisions. Accurate determination of FFs is essential for testing QCD factorization, 
deepening our understanding of hadronization mechanisms, and refining global fits aimed at extracting parton distribution functions (PDFs). 
Traditionally, SIA data have provided the primary constraints on FFs due to their straightforward theoretical interpretation. 
However, SIDIS measurements are indispensable for distinguishing quark and antiquark fragmentation and 
achieving comprehensive flavor separation~\cite{AbdulKhalek:2022laj,Gao:2024dbv,Soleymaninia:2024jam}. 

The Electron-Ion Collider (EIC), scheduled to commence operations around 2030 at Brookhaven National Laboratory, 
will be a state-of-the-art, high-luminosity facility designed to systematically investigate the internal structure 
of nucleons and nuclei with unprecedented precision~\cite{AbdulKhalek:2019mzd,Klasen:2018gtb,Klasen:2017kwb,AbdulKhalek:2022hcn,AbdulKhalek:2021gbh}. 
This new capability promises to significantly advance 
our comprehension of QCD dynamics, particularly illuminating the interplay between sea quarks and gluons~\cite{AbdulKhalek:2021gbh,Radici:2022hbs}.

One of the main scientific objectives of the EIC will be high-precision DIS and SIDIS measurements, providing 
detailed and differential constraints on both PDFs and FFs~\cite{Zhou:2021llj,Khalek:2021ulf,Armesto:2023hnw,Jimenez-Lopez:2024hpj}. In addition to 
improving the determination of sea quark and gluon distributions, simulated DIS data 
from the EIC will also enable enhanced constraints on valence quark distributions. Recent studies indicate that 
the inclusion of these projected datasets significantly reduces uncertainties in the strong coupling constant 
extraction, driven by the substantial extension of the accessible kinematic ranges in both Bjorken $x$ and $Q^2$~\cite{Kutz:2024eaq,Azizi:2024swj,Cerci:2023uhu}. 
Furthermore, the high-statistics SIDIS data expected from the EIC will substantially decrease uncertainties in 
FF extractions, especially for charge-separated quark distributions and gluon fragmentation. Thus, EIC measurements will 
complement and enhance existing constraints from SIA and other high-energy processes, facilitating significant refinements 
in global determinations of FFs. 

The impact of projected SIDIS EIC data on proton PDFs as well as pion and kaon fragmentation functions was 
previously investigated in Ref.~\cite{Aschenauer:2019kzf}, employing a reweighting technique to incorporate 
realistic pseudo-data uncertainties into established global extractions. Their analysis utilized cross-section 
data unfolded for detector effects, highlighting the substantial improvements expected for charged pion and kaon FFs. 

In this study, we perform a comprehensive global QCD analysis integrating charged pion SIDIS pseudo-data from the 
EIC with all available SIA and real SIDIS datasets. Our analysis represents the first instance of directly 
incorporating EIC SIDIS pseudo-data into a global fit to quantify their precise impact on FF extractions. 
A significant feature of our approach is the adoption of a neural network-based methodology, which provides enhanced 
flexibility and accuracy compared to traditional parametrization techniques.

Our findings demonstrate that the inclusion of EIC pseudo-data substantially improves the precision of pion FFs, 
particularly at the previously poorly constrained large-momentum fraction ($z$) region. 
In fact, uncertainty bands of FFs are notably reduced, particularly for 
gluon and up-quark FFs when EIC pseudo-data are included. This underscores the critical role future EIC data will play in addressing the persistent challenge of 
precisely constraining gluon FFs. 

The paper is structured as follows: Section~\ref{sec:theory} outlines the theoretical framework for inclusive hadron 
production in electron-positron annihilation and semi-inclusive deep-inelastic scattering, 
describes our neural network-based parametrization of FFs, and details the methodology for uncertainty 
estimation. Section~\ref{sec:data} introduces the experimental datasets used, including a discussion of the 
generation of EIC pseudo-data and applied kinematic cuts. Section~\ref{results} presents our main results, 
focusing on the quality of the global fit and the specific impacts of incorporating EIC data on pion FFs, along 
with a detailed comparison with previous results from the MAPFF and NNFF1.0 analyses. Finally, Section~\ref{summary} 
summarizes our conclusions and outlines potential future directions for extending this analysis.

\section{Theoretical and Methodological Setup}\label{sec:theory}

In this analysis, we investigate charged pion FFs utilizing experimental data from two complementary processes: the SIA and SIDIS. 
Each of these processes offers distinct and essential constraints on the flavor-dependent FFs. 
The cross-section for the SIA process, describing hadron production in electron-positron annihilation, 
is expressed as a convolution of a perturbatively calculable hard-scattering 
coefficient function and non-perturbative FFs~\cite{Soleymaninia:2024jam,Moffat:2021dji}: 

\begin{equation}
\frac{d\sigma^{\text{SIA}}}{dQ^2 dz_h} = \sum_j C_j^{\text{SIA}}(z_h, Q^2) \otimes D_j(z_h, \mu),
\end{equation}

where \( z_h \) represents the fraction of momentum carried by the observed hadron 
relative to the parent parton, and \( \otimes \) denotes the convolution integral.

Similarly, the SIDIS process involves an electron scattering off a nucleon with the subsequent detection of a hadron in the final state. 
The SIDIS cross-section is given by a double convolution involving the PDFs, FFs, and the corresponding hard-scattering coefficient functions:

\begin{equation}
\frac{d\sigma^{\text{SIDIS}}}{dQ^2 dx_B dz_h} = \sum_{i,j} C_{ij}^{\text{SIDIS}}(x_B, z_h, Q^2) \otimes f_i(x_B, \mu) \otimes D_j(z_h, \mu).
\end{equation}

In this expression, \( x_B \) is the Bjorken scaling variable, and again \( z_h \) is the hadronic momentum fraction. 
Additional theoretical details and derivations can be found in Refs.~\cite{Soleymaninia:2024jam,Moffat:2021dji}.

Our global analysis of SIA and SIDIS data is carried out within a perturbative QCD framework at NLO accuracy. 
To parametrize the FFs, we adopt a flexible Neural Network (NN) approach, enabling robust and unbiased uncertainty propagation 
from both experimental data and input PDFs. Specifically, the NNPDF4.0 PDF sets~\cite{NNPDF:2021njg} 
are employed in calculating the SIDIS cross-sections. 
To perform this global QCD analysis, we utilize the 
publicly available {\tt MontBlanc} package~\cite{Khalek:2021gxf,AbdulKhalek:2022laj}, accessible from~\cite{MontBlanc}. 

As heavy-quark mass corrections for SIDIS processes are not yet well-established, we perform our analysis 
within the Zero-Mass Variable Flavor Number Scheme (ZM-VFNS). 
In this approximation, all active partons are treated as massless; however, we incorporate partial heavy-quark mass 
dependence by matching different active-flavor number sub-schemes at heavy-quark thresholds. 
The charm and bottom thresholds are set at \( m_c = 1.51 \, \text{GeV} \) and \( m_b = 4.92 \, \text{GeV} \), respectively, 
consistent with the NNPDF4.0 sets~\cite{NNPDF:2021njg,NNPDF:2024nan}. 
Below these thresholds, heavy-quark FFs remain constant and do not evolve. 

In this analysis, we utilize a framework that combines a neural network parametrization of FFs with a 
Monte Carlo representation of FF uncertainties. This approach, which has been extensively used by 
the NNPDF Collaboration to determine PDFs of the proton and of nuclei, allows us to 
reduce model bias in FF parametrization as much as possible, and to faithfully propagate experimental and 
PDF uncertainties into FFs~\cite{AbdulKhalek:2022laj, Khalek:2021gxf}. 
These features are essential to achieve the methodological accuracy of FFs that are utilized to analyze, e.g., 
high-precision hadron production measurements at the Large Hadron Collider (LHC) and, in the future, at the EIC.

The neural network is optimized by means of a gradient descent algorithm that makes use of the knowledge of 
the analytic derivatives of the neural network itself with respect to its free parameters~\cite{AbdulKhalek:2022laj}.

As discussed in our previous work~\cite{Soleymaninia:2022alt}, the inclusive SIA data only allow for the 
determination of the total summed quark and antiquark FFs by including the total 
inclusive cross-sections for light, charm, and bottom quarks.  
By incorporating SIDIS data sets into the analysis, we gain the ability to directly constrain the 
individual quark and antiquark FFs for light quarks. For the parametrization of \( \pi^+ \), we adopt the following basis:

\begin{eqnarray}
\label{eq:combinations}
D^{\pi^+}_{u}, ~D^{\pi^+}_{\bar{d}}, ~D^{\pi^+}_{d}=D^{\pi^+}_{\bar{u}}, 
~D^{\pi^+}_{s^+}, ~ D^{\pi^+}_{c^+}, 
~D^{\pi^+}_{b^+}, ~ D^{\pi^+}_g\,.
\end{eqnarray}

The parameterization is introduced at the initial scale \( \mu_0 = 5 \, \text{GeV} \). 
To reduce model bias in the parametrization of FFs, we employ a feed-forward neural network (NN) architecture at the initial scale. The FFs for each parton flavor are defined through the NN model as
\begin{eqnarray}
	\label{NN}
	zD^{h^+}_i(z,Q_0)
	= 
	(N_i(z; \boldsymbol {\theta}) - N_i(1; \boldsymbol {\theta}))^2 
	\, ,
\end{eqnarray}
where $N_i(z;\boldsymbol\theta)$ is the NN output and $\theta$ stands for the (internal) parameters of the NN. The subtraction at $z=1$ ensures that the FFs vanish at the endpoint, and squaring the result guarantees positivity. We deliberately avoid introducing pre-factors such as $z^\alpha(1-z)^\beta$ to regulate the endpoint behavior, thereby minimizing theoretical bias and allowing the data to drive the shape of the FFs.

The neural network is a multilayer feed-forward perceptron~\cite{Forte:2002fg}. The output $\xi_i^{(l)}$ of the $i$-th node in the $l$-th layer is given by
\begin{equation}
	\xi_i^{(l)}=g\left(\sum_j\omega_{ij}^{(l)}\xi_j^{(l-1)}+\theta_i^{(l)} \right)
	\, ,
	\label{eq:actfunc}
\end{equation}
where $g$ is the activation function (sigmoid in hidden layers, linear in the output layer), and $\omega_{ij}^{(l)}$ and $\theta_i^{(l)}$ are the weights and biases, respectively. These parameters are determined by minimizing the $\chi^2$ function.

We use a (1-20-7) architecture: one input node (for $z$), one hidden layer with 20 nodes, and seven output nodes for the FF combinations. While the use of NNs can mitigate model bias, we acknowledge that design choices such as activation functions and constraints can affect uncertainty estimates. This effect is discussed, for example, in NNPDF4.0~\cite{NNPDF:2021njg}. Similar strategies to ours are also adopted in recent global FF fits such as Ref.~\cite{AbdulKhalek:2022laj,Soleymaninia:2023dds}, where the flexibility of the NN allows for unbiased extraction of FFs even in the underconstrained small- and large-$z$ regions.

We choose the initial scale of parametrization for the fragmentation functions to be $Q_0 = 5~\mathrm{GeV}$. This scale lies above the bottom-quark mass threshold ($m_b = 4.92~\mathrm{GeV}$). This choice is advantageous for two reasons: first, it avoids any heavy-quark threshold crossings during evolution, keeping the number of active flavors fixed at $n_f = 5$, and thereby eliminates the need for time-like matching conditions across thresholds, which are currently only known to NLO accuracy \cite{Cacciari:2005ry}. Second, in the variable flavor number scheme (VFNS), this allows us to parametrize charm and bottom FFs on the same footing as light-quark and gluon FFs, accommodating their significant non-perturbative components. Although some SIDIS data lie below $Q_0$, we perform backward evolution using numerically stable time-like DGLAP evolution routines, which have been validated down to $Q \sim 1.4~\mathrm{GeV}$. This approach is consistent with previous global analyses such as MAPFF~\cite{AbdulKhalek:2022laj, Khalek:2021gxf}, where the same initial scale is adopted and reliable backward evolution is achieved. We confirm that the numerical stability of our evolution routines is preserved throughout the full $Q^2$ range of our analysis.

By including the SIDIS data in the QCD fit, the combinations 
of quark FFs, \( D^{\pi^{\pm}}_{u^+} \) and \( D^{\pi^{\pm}}_{d^+} \), are considered to be decomposed.  
The strange quark and heavy quark distributions are assumed to be symmetric, as expressed by:

\begin{eqnarray}
\label{eq:Symmetry}
 D^{\pi^+}_q = D^{\pi^+}_{\bar{q}},~~~~~~~~ q = s, c, b\,.
\end{eqnarray}

Thus, by adding the SIDIS data, the number of independent distributions increases to seven in comparison 
to five in \cite{Bertone:2017tyb}. We find that these flavor combinations for 
quark FFs provide the best fit quality and accuracy in hadron production.

Uncertainties on the FFs are determined by means of the Monte Carlo sampling method properly accounting for 
all sources of experimental uncertainties, including that of parton distribution functions. 
This method allows us to incorporate the uncertainty information coming from datasets and input PDFs using Monte Carlo, avoiding 
a full time-consuming refit, but preserving the statistical rigor for the uncertainty estimates \cite{Khalek:2021gxf}.

In the present work, hadron mass and target mass corrections (TMCs) have not been included in our analysis. 
We intend to address these aspects in future studies. Additionally, forthcoming analyses will integrate recent 
experimental data, such as the BESIII measurements for \(\pi^\pm\) production in 
electron-positron annihilation~\cite{BESIII:2025mbc}, as well as newly reported charged pion multiplicities 
from SIDIS measurements on proton and deuteron targets provided by the JLab collaboration~\cite{Bhatt:2024prq}.

\section{Experimental Data}\label{sec:data}

This section details the experimental data sets analyzed in our work, including the SIA and SIDIS, highlighting their roles in 
constraining FFs and specifying their kinematic coverage. 
It also introduces and describes the simulated EIC pseudo-data for SIDIS measurements, emphasizing their 
complementary role and discussing the covered kinematic ranges. 
Finally, this section outlines the criteria used for data selection and describes the specific kinematic cuts applied in our 
analysis to ensure theoretical consistency and reliability.

%
\subsection{SIA and SIDIS data sets}\label{SIA-SIDIS}
%

SIA measurements directly probe the FFs integrated over quark and antiquark flavors. 
Due to their inclusive nature, these measurements primarily constrain the charge-weighted sum of FFs, limiting sensitivity to 
individual flavor distinctions. However, the availability of SIA data sets from different experiments at 
multiple center-of-mass energies allow for constraints on the gluon FF 
through Dokshitzer-Gribov-Lipatov-Altarelli-Parisi (DGLAP) evolution~\cite{Dokshitzer:1977sg,Gribov:1972ri,Altarelli:1977zs}. 
In contrast, SIDIS provides direct sensitivity to individual quark and antiquark FFs due to the flavor-dependent 
nature of PDFs. SIDIS measurements typically cover lower energy scales ($1 \leq Q^2 \leq 8$ GeV$^2$), significantly aiding the 
determination of gluon FF via QCD evolution. Moreover, experiments using isoscalar targets such as deuterium and 
lithium further enhance flavor separation, particularly distinguishing valence from sea quark contributions.
Therefore, the combination of both SIA and SIDIS datasets provides a powerful, complementary approach to FF extraction. 
This complementarity, reinforced by carefully chosen kinematic cuts, ensures theoretical consistency and robustness 
in the global extraction of FFs. 
Details on the specific experimental datasets used from SIA and SIDIS processes are provided in the following subsections.

%
\subsubsection{The SIA data sets}  
%

SIA provides a clean experimental environment for studying fragmentation, 
as it does not involve the complexities of initial-state parton distributions. In this process, 
an electron and a positron annihilate into a virtual photon (or Z boson), 
which subsequently produces a quark-antiquark pair that fragments into final-state hadrons 

\begin{equation}
e^+ + e^- \to h + X.
\end{equation}

The experimental SIA data used in this analysis are collected from various high-energy 
experiments covering a wide range of center-of-mass energies \( \sqrt{s} \):  

\begin{itemize}
\item \textbf{CERN:} ALEPH~\cite{ALEPH:1994cbg}, DELPHI~\cite{DELPHI:1998cgx}, OPAL~\cite{OPAL:1994zan}
\item \textbf{DESY:} TASSO~\cite{TASSO:1980dyh, TASSO:1982bkc, TASSO:1988jma}
\item \textbf{KEK:} BELLE~\cite{Belle:2013lfg}, TOPAZ~\cite{TOPAZ:1994voc}
\item \textbf{SLAC:} BABAR~\cite{BaBar:2013yrg}, TPC~\cite{TPCTwoGamma:1988yjh}, SLD~\cite{SLD:2003ogn}
\end{itemize}

The measurements span a kinematic range from \( \sqrt{s} \sim 10 \) GeV (B-factory experiments such as BABAR and BELLE) to 
the  \( \sqrt{s} = M_Z \) (LEP experiments such as ALEPH, DELPHI, OPAL, and SLD).

%
\subsubsection{The SIDIS data sets}  
%

In the SIDIS process, a charged lepton scatters off a nucleon, exchanging a virtual photon that probes the 
nucleon's parton structure. A hadron is detected in the final state, leading to the reaction  

\begin{equation}
\ell + N \to \ell' + h + X.
\end{equation}

The cross-section depends on both the partonic structure of the nucleon (PDFs) and the hadronization of the struck quark.   

The SIDIS data used in this study are collected from:  

\begin{itemize}
\item \textbf{COMPASS} (CERN): Muon beam (\( E_\mu = 160 \) GeV) on a $^6$LiD target~\cite{COMPASS:2016xvm}.
\item \textbf{HERMES} (DESY): Electron/positron beams (\( E_e = 27.6 \) GeV) on hydrogen and deuterium targets~\cite{HERMES:2012uyd}.
\end{itemize}

%
\subsection{Simulated EIC SIDIS data}\label{EIC}
%

In addition to the SIA measurements from LEP, DESY, SLAC, and KEK, and SIDIS data from HERMES and COMPASS, this analysis 
incorporates the EIC pseudo-data for the semi-inclusive deep-inelastic scattering process. 

We show that the projected EIC measurements for SIDIS at $\sqrt{s} = 140$ GeV and $\sqrt{s} = 45$ GeV provide crucial constraints on FFs. 
These datasets are generated using Monte Carlo simulations with realistic experimental uncertainties, 
accounting for detector effects and kinematic acceptance criteria. 
The EIC is designed as a high-energy, high-luminosity polarized collider ($\mathcal{L} \sim 10^{33} - 10^{34}$ cm$^{-2}$ s$^{-1}$) and 
has evolved through two preconceptual designs.

To span the full kinematic coverage of the EIC, studies are performed for lepton beam energies of 5 GeV and 20 GeV, 
in combination with proton beam energies of 100 GeV and 250 GeV, respectively. 
The pseudo-data are generated using the Monte Carlo event generator PYTHIA-6 to simulate DIS events, 
assuming a statistical uncertainty corresponding to an integrated luminosity of 10 fb$^{-1}$. 
Only events satisfying $Q^2 > 1$ GeV$^2$, a squared invariant photon-nucleon mass of $W^2 > 10$ GeV$^2$, and an 
inelasticity cut of $0.01 < y < 0.95$ are considered.
For further details about EIC SIDIS data, we refer the reader to Ref.~\cite{Aschenauer:2019kzf}.

The two energy configurations complement each other in kinematic coverage: the $\sqrt{s} = 140$ GeV data extend the reach to lower $x_B$ values, 
probing the sea-quark region, whereas the $\sqrt{s} = 45$ GeV data provide complementary constraints at higher $x_B$, aiding in the extraction of 
valence-quark distributions. The inclusion of these pseudo-data in global fits is 
expected to significantly reduce uncertainties, particularly for the gluon FF at medium to large $z$ and for the charge-separated light-quark FFs. 
These improvements highlight the EIC transformative role in refining our understanding of fragmentation processes and hadronization mechanisms in QCD.

We emphasize that the pseudo--data included in our analysis are not generated using general-purpose event generators such as \textsc{Pythia6}, which are based on LO matrix elements, parton showers, and phenomenological hadronization models. Instead, we adopt a fully perturbative approach based on QCD collinear factorization at NLO. Specifically, the central values of the SIDIS cross sections are computed using the \texttt{MontBlanc} program, which performs NLO calculations via fast convolutions of stored grids of hard coefficient functions with input fragmentation functions \texttt{MAPFF} \cite{AbdulKhalek:2022laj}. This ensures consistency with the theoretical framework used in our global QCD analysis, which employs two-loop time-like DGLAP splitting kernels for FF evolution.

The pseudo--data generation follows a controlled procedure: each bin's central value \(\sigma_i^{\mathrm{exp}}\) is derived from the theoretical prediction \(\sigma_i^{\mathrm{theory}}\) as
\begin{align}
	\sigma_i^{\mathrm{exp}} = \sigma_i^{\mathrm{theory}} \left( 1 + r_i \, \delta_{i}^{\mathrm{uncorr}} + s \, \delta^{\mathrm{corr}} \right),
	\label{eq:pseudodataGen}
\end{align}
where \(r_i\) and \(s\) are independent Gaussian random variables, \(\delta_i^{\mathrm{uncorr}}\) is the uncorrelated uncertainty in bin \(i\), and \(\delta^{\mathrm{corr}}\) is the correlated systematic uncertainty, set to \(3.5\%\) across all bins. The uncorrelated uncertainty is defined as
\begin{align}
	\delta_i^{\mathrm{uncorr}} = \sqrt{ \left( \delta_i^{\mathrm{stat}} \right)^2 + \left( \delta_i^{\mathrm{sys}} \right)^2 }.
	\label{eq:totalExpError}
\end{align}

Subsequently, the central values are subjected to random fluctuations according to the above formula. By construction, this ensures that the calculated 
	\(\chi^2/N_{\text{data}}\) for the pseudo--data is expected to be close to unity, thereby validating the internal consistency between the generated data and the theoretical model.

With this procedure, the pseudo--data reflect the expected statistical and systematic precision of future lepton colliders, while being fully compatible with the NLO theoretical framework used in our analysis. Since the pseudo--data are not generated using LO parton shower models or hadronization schemes, no theoretical mismatch is introduced, and their inclusion does not create tension with existing SIDIS or SIA measurements. This justifies their role in evaluating the constraining power of future EIC measurements on fragmentation functions.

\begin{figure}[t]
	\centering
	\includegraphics[width=0.5\textwidth]{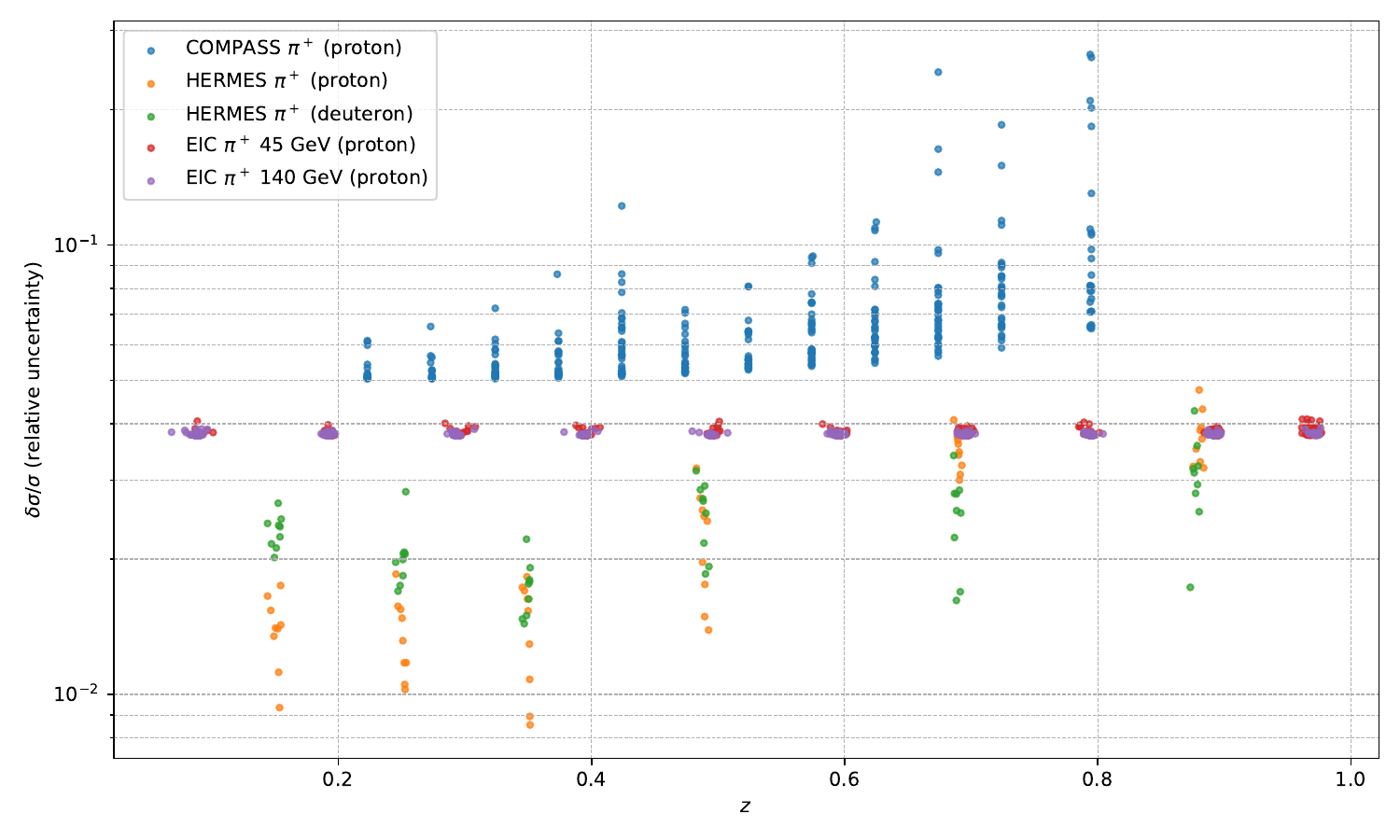}
	\caption{Relative total uncertainties $\delta\sigma/\sigma$ versus $z$ for the $\pi^+$ SIDIS data sets: HERMES, COMPASS, and EIC ($\sqrt{s}$ = 45 and 140 GeV).}
	\label{fig:relerr-piplus}
\end{figure}

\begin{figure}[t]
	\centering
	\includegraphics[width=0.5\textwidth]{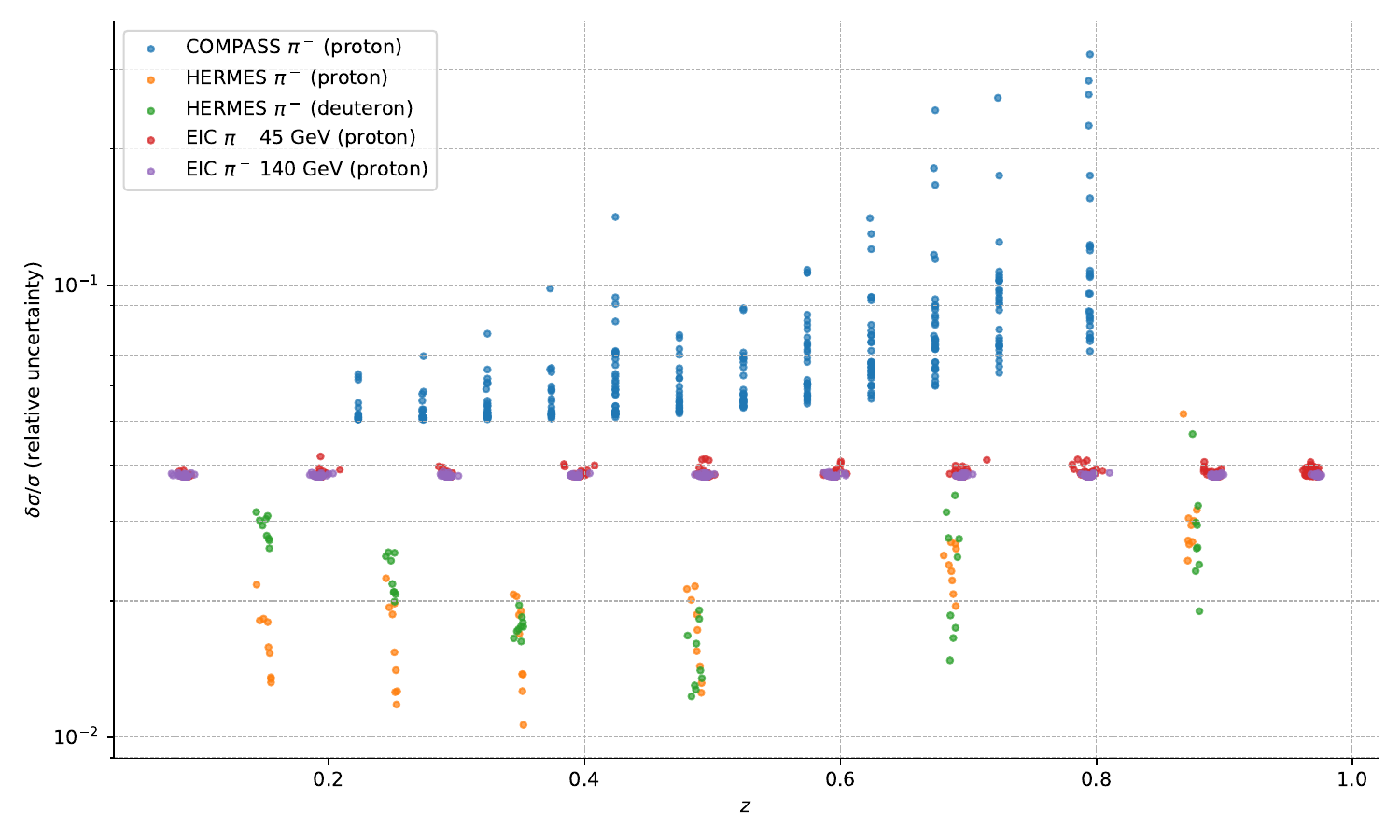}
	\caption{Same as Fig.~\ref{fig:relerr-piplus}, but for the $\pi^-$ SIDIS data sets.}
	\label{fig:relerr-piminus}
\end{figure}

To further clarify the relative precision of the simulated EIC pseudo--data compared to existing measurements, we present in Figs.~\ref{fig:relerr-piplus} and \ref{fig:relerr-piminus} the relative total uncertainties $\delta\sigma/\sigma$ as a function of the fragmentation variable $z$ for the $\pi^+$ and $\pi^-$ multiplicity datasets, respectively. These figures include the SIDIS data from HERMES (proton and deuteron targets), COMPASS, and the projected EIC pseudo--data at $\sqrt{s} = 45$~GeV and $140$~GeV. It is clearly observed that the EIC pseudo--data exhibit significantly smaller uncertainties than COMPASS data at all range of $z$ for both $\pi ^+$ and $\pi ^-$. 
However, HERMES measurements typically have smaller uncertainties across a broad $z$ range in comparison to EIC pseudo--data, particularly at low values of $z$. This highlights the enhanced constraining power of the EIC data and supports the conclusion that their inclusion can lead to substantially improved determinations of fragmentation functions in global QCD analyses.

\subsection{Data Selection and Kinematical Cuts}\label{sec:cuts}

\begin{figure}[!htb]
\begin{center}
\includegraphics[scale = 0.55]{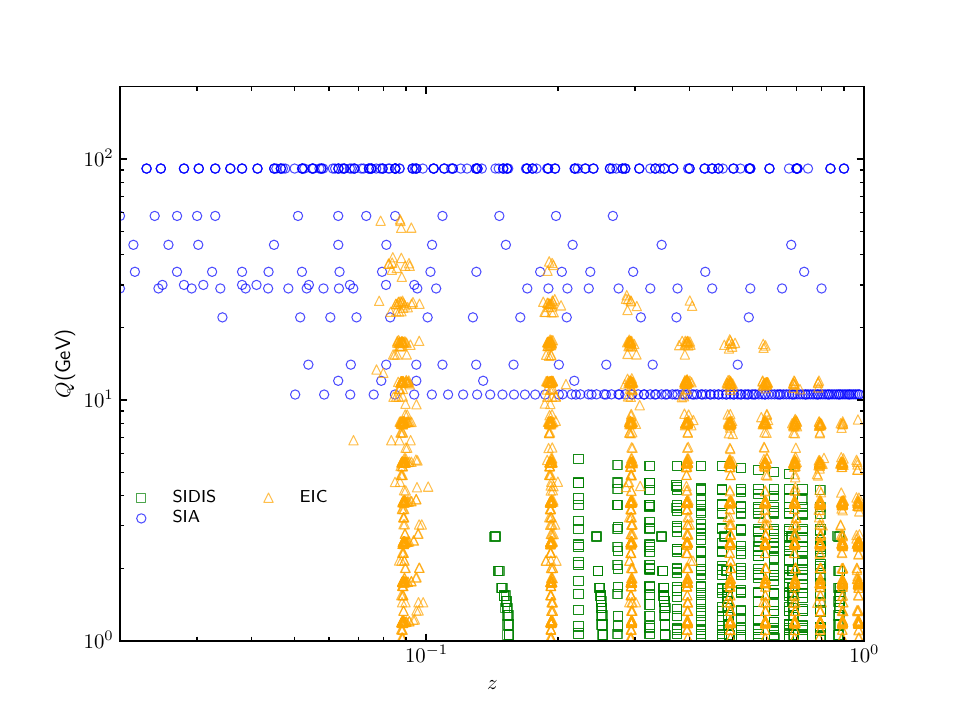}
\caption{Kinematic coverage in the $(z, Q)$ plane for the SIDIS, SIA, and EIC simulated SIDIS pseudo-data included in this analysis.}
\label{fig:Data}
\end{center}
\end{figure}

In this analysis, we incorporate all available experimental datasets on pion production from SIA and SIDIS experiments, 
along with the simulated SIDIS pseudo-data from the EIC, to extract pion FFs. 
Below, we discuss the kinematic coverage of these datasets and detail the specific cuts on the 
fragmentation variable $z$ and the energy scale $Q$ applied throughout this study.
Figure~\ref{fig:Data} shows the coverage of the SIDIS, SIA, and EIC datasets in the $(z, Q)$ plane. 

The SIA data predominantly cover higher $Q$ regions with an extended range in $z$. 
In contrast, the SIDIS data span a wide range of $z$, covering low to intermediate values and partially extending into the high-$z$ region. 
The moderate range of $Q$ covered by SIDIS provides valuable flavor-sensitive information on 
fragmentation across different energy scales. 
The projected EIC measurements substantially enhance the existing kinematic coverage by extending into higher $Q$ 
regions while maintaining a comprehensive span in $z$. This extended coverage is especially beneficial for the high-$z$ region, 
where current FF extractions suffer from larger uncertainties.  
By bridging SIDIS and SIA data, the EIC pseudo-data significantly improve constraints on both quark and gluon FFs, 
enhance flavor separation, and refine our understanding of hadronization mechanisms across diverse kinematic domains.

To ensure the validity of perturbative QCD and the applicability of QCD evolution, we impose specific 
kinematic cuts on the data. Particularly for the SIA data, we exclude very low values of $z$, as this region is dominated 
by non-perturbative effects. Following established analyses~\cite{AbdulKhalek:2022laj,Soleymaninia:2022alt}, we restrict 
the SIA data to the kinematic range $0.02 \leq z \leq 0.9$ for center-of-mass energies near 
the $Z$-boson mass ($M_Z$), and to $0.075 \leq z \leq 0.9$ for other center-of-mass energies. 
For the SIDIS data from HERMES and COMPASS, we retain data points within $0.2 \leq z \leq 0.8$. The broader and more precise 
kinematic reach of the EIC pseudo-data allows us to extend the $z$ coverage to $0.1 \leq z \leq 0.98$, 
enabling a more comprehensive extraction of FFs across an extended phase space. 

At low energy scales ($Q$), higher-order perturbative corrections become increasingly significant, 
making it necessary to account for them to maintain reliable theoretical predictions. 
However, at NLO accuracy, perturbative corrections become less reliable at very low $Q$. 
Therefore, we exclude data at very small $Q$ values from the SIDIS datasets to ensure the robustness and reliability of our extractions.

\begin{figure}[!htb]
\begin{center}
\includegraphics[scale = 0.450]{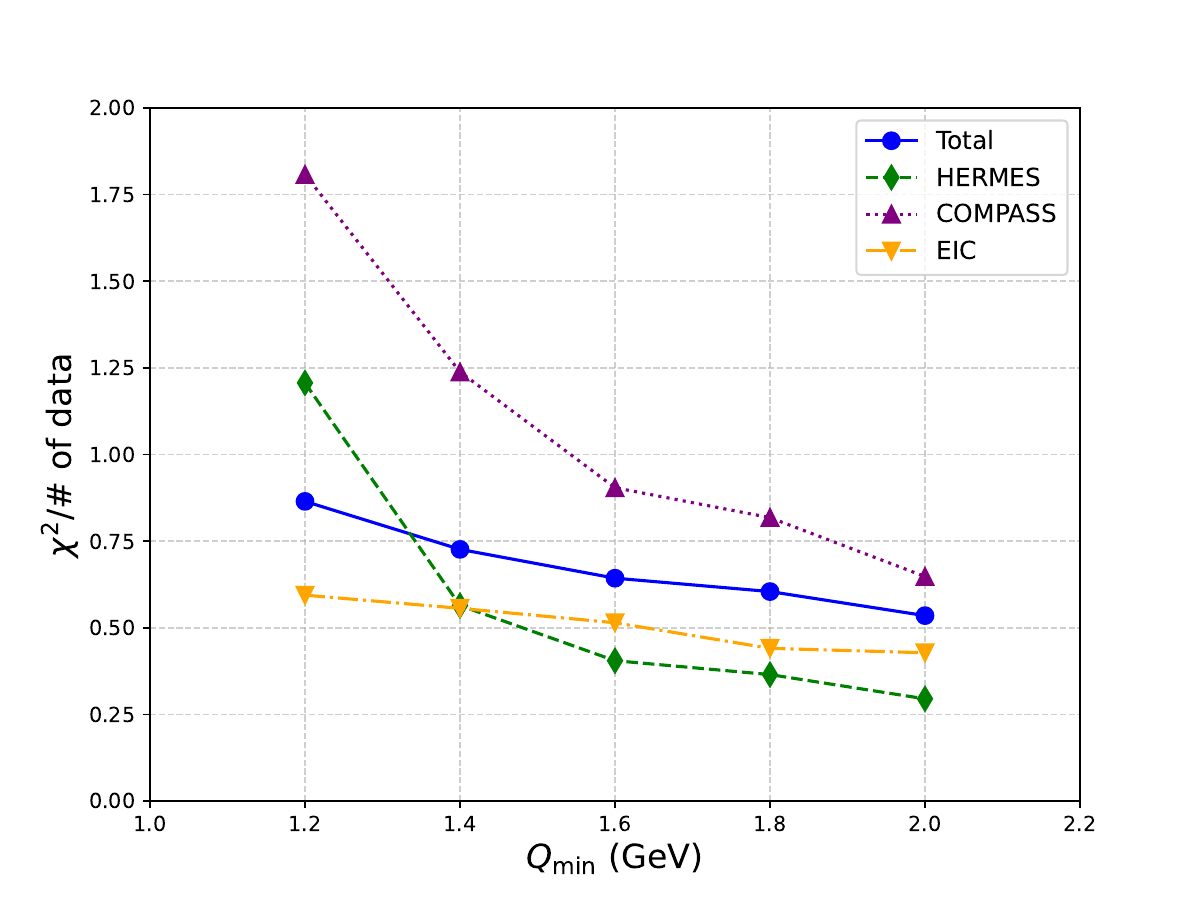}
\caption{Variation of $\chi^2$ per data point as a function of the applied $Q_{\text{min}}$ cut in the SIDIS datasets. 
The curves show $\chi^2/N_{\text{data}}$ for the total dataset (blue), HERMES (green), COMPASS (purple), 
and EIC pseudo-data (yellow) at different $Q_{\text{min}}$ thresholds.} 
\label{fig:chi2_scanning}
\end{center}
\end{figure}

To determine an appropriate lower bound on $Q$, we studied how the fit quality depends on the choice of $Q_{\text{min}}$. 
Figure~\ref{fig:chi2_scanning} shows the variation of $\chi^2/N_{\text{data}}$ as a function of the 
minimum energy scale $Q_{\text{min}}$ (in GeV) for different SIDIS datasets used in the extraction of pion FFs at NLO accuracy.
The trend of the total $\chi^2/N_{\text{data}}$ reflects the interplay among datasets and the constraints they 
impose for different choices of $Q_{\text{min}}$.

The SIDIS datasets from HERMES and COMPASS show increasing values of $\chi^2/N_{\text{data}}$ 
at lower $Q_{\text{min}}$, indicating rising theoretical tensions and uncertainties at these lower energy scales. 
This suggests that SIDIS data at low $Q$ introduce greater variability in the fit, 
potentially due to non-perturbative contributions or limitations inherent in the NLO pQCD framework. 
In contrast, the inclusion of EIC pseudo-data has a stabilizing effect, particularly noticeable at small $Q_{\text{min}}$, 
mitigating fluctuations and significantly improving overall fit consistency. 
This highlights the crucial role of the EIC pseudo-data in providing additional constraints 
and yielding a more stable and reliable determination of FFs at lower energy scales.

Based on the observed behavior of $\chi^2/N_{\text{data}}$, we choose $Q_{\text{min}} = 1.4$ GeV as the optimal lower cutoff. 
This selection ensures an appropriate balance between retaining sufficient data at low energy scales and minimizing the tensions indicated  
by higher $\chi^2/N_{\text{data}}$ values observed at lower $Q_{\text{min}}$. 
The stability of the fit quality around this chosen value indicates that it represents the 
optimal compromise between precision and consistency in the global fit. 
After applying these kinematic cuts, our final dataset includes a total of $N_{\text{data}} = 2434$ data points, 
comprising $N_{\text{data}} = 498$ from SIDIS, $N_{\text{data}} = 1559$ from 
EIC pseudo-data, and $N_{\text{data}} = 377$ from SIA measurements. 

In summary, our results confirm the complementary nature of the different datasets. The  HERMES and COMPASS SIDIS data provide essential information at 
moderate energy scales, while EIC pseudo-data significantly reduce uncertainties and enhance precision, 
particularly in lower energy regions.  
Our results show that the 
inclusion of EIC pseudo-data leads to a more constrained, stable, and consistent extraction of FFs, emphasizing 
its importance for future high-precision QCD analyses.

%
\section{Results}\label{results}
%

In this section, we present the main results and key findings of our global QCD analysis. 
We conducted two distinct QCD fits: the first fit, labeled ``Global Fit,'' incorporates the SIA data along 
with SIDIS measurements from the HERMES and COMPASS experiments, while the second fit, 
labeled ``Global Fit + EIC,'' additionally includes simulated SIDIS EIC data.
We begin by evaluating the quality of these fits by examining the $\chi^2/N_{\text{data}}$ values 
obtained for each experimental dataset considered. Next, we present the extracted 
pion FFs resulting from the combined analysis of SIA, SIDIS, and EIC datasets. 
We then assess the specific impact of incorporating projected EIC measurements on the precision 
of the FFs for different parton species. 
Finally, we compare our extracted FFs to other existing sets available in the literature.

%
\subsection{Assessment of fit quality and dataset consistency}\label{fit-quality} 
%

\begin{figure}[!htb]
\begin{center}
\includegraphics[scale = 0.35]{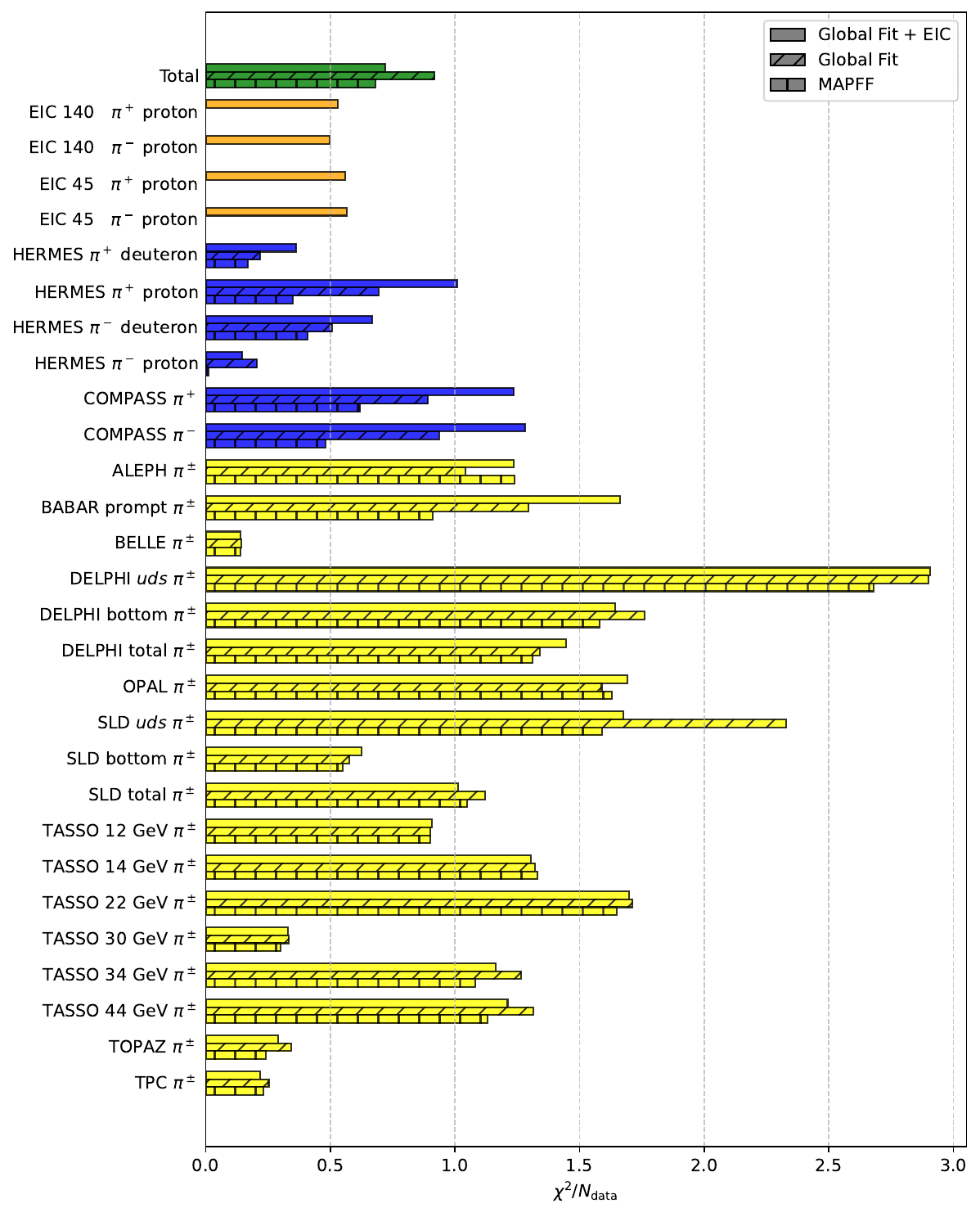}
\caption{Comparison of the \( \chi^2 / N_{\text{data}} \) values for three fits: Global Fit, Global Fit + EIC and MAPFF.  
The yellow bars represent the SIA datasets, the blue bars denote the SIDIS datasets and the orange bars show the EIC datasets. 
The topmost green bar summarizes the total \( \chi^2 / N_{\text{data}} \) across all datasets combined.
} 
\label{fig:chi2}
\end{center}
\end{figure}

We discuss the quality of these fits by examining the $\chi^2/N_{\text{data}}$ value obtained for each experimental dataset for both QCD fits.
Figure~\ref{fig:chi2} shows the values of $\chi^2/N_{\text{data}}$ for the various experimental datasets 
included in our analysis, categorized into SIDIS, SIA, and EIC pseudo-data. 
The SIDIS measurements are from the HERMES and COMPASS experiments, while the SIA datasets include results from 
multiple electron-positron annihilation experiments, specifically BELLE, BABAR, TASSO, TPC, TOPAZ, ALEPH, DELPHI, OPAL, and SLD. 
As previously discussed, the EIC pseudo-data incorporated into this analysis correspond to projected 
measurements at center-of-mass energies of 45 GeV and 140 GeV.

The distribution of $\chi^2/N_{\text{data}}$ values provides insights into the consistency of each 
dataset within the global fit. For the SIDIS datasets (HERMES and COMPASS), the observed moderate $\chi^2$ values reflect 
the complexity associated with modeling hadronic final states in SIDIS processes. 
Among the SIA datasets, lower $\chi^2$ values for experiments such as BELLE, TOPAz and TPC highlight their robust 
and stable constraints derived from clean $e^+e^-$ annihilation data. In contrast, certain datasets, notably the DELPHI $uds$ measurements, 
exhibit higher $\chi^2/N_{\text{data}}$ values, likely due to greater systematic uncertainties or mild internal tensions. 
This observation is consistent with previous results from the MAPFF analysis~\cite{AbdulKhalek:2022laj}.

The EIC data particularly improve precision in kinematic regions where current SIDIS measurements provide limited accuracy. The total $\chi^2/N_{\text{data}}$, represented by the green bars in Figure~\ref{fig:chi2}, summarizes the overall quality 
of the global fit across all datasets. 
The $\chi^2/N_{\text{data}}$ for the Global Fit is calculated as 0.919, and for the Global Fit + EIC, it is 0.721. 
These values suggests that our global analysis 
successfully balances the constraints from diverse experimental sources, effectively accommodating systematic 
uncertainties and variations inherent in the individual datasets.

In Fig.~\ref{fig:chi2}, the \(\chi^2/N_{\text{data}}\) values are also compared with the MAPFF results~\cite{AbdulKhalek:2022laj} for all individual data sets. Both our ``Global Fit'' analysis and the MAPFF analysis utilize the same data sets, with minor differences in the kinematic cuts applied to the HERMES and COMPASS SIDIS data. Specifically, MAPFF applies a virtuality cut of \(Q_{\text{cut}} = 2~\text{GeV}\) for the HERMES and COMPASS SIDIS data sets, whereas our analysis uses a looser cut of \(Q_{\text{cut}} = 1.4~\text{GeV}\) for all SIDIS data. In addition, MAPFF employs the NNPDF3.1~\cite{NNPDF:2017mvq} PDF sets for computing SIDIS cross sections, while we use the more recent NNPDF4.0~\cite{NNPDF:2024dpb} sets.
Although both analyses are based on the MontBlanc framework, some differences in the results can be observed. These differences primarily arise from the variation in the number of data points included from each data set, as well as from the choice of PDF sets used in the two analyses.

%
\subsection{Impact of EIC pseudo-data on the precision of Pion FFs}\label{Impact}  
%

The anticipated SIDIS measurements from the EIC are expected to significantly affect the global extraction of FFs. 
These measurements are crucial for global QCD analyses, as SIDIS data primarily facilitate the separation of 
quark and antiquark FFs, substantially enhancing flavor differentiation. 
To quantitatively assess the impact of future EIC SIDIS measurements, 
we performed two sets of QCD fits with and without EIC SIDIS pseudo-data. 

\begin{figure*}[htb]
\vspace{0.0010cm}
\centering
\subfloat{\includegraphics[width=0.39\textwidth]{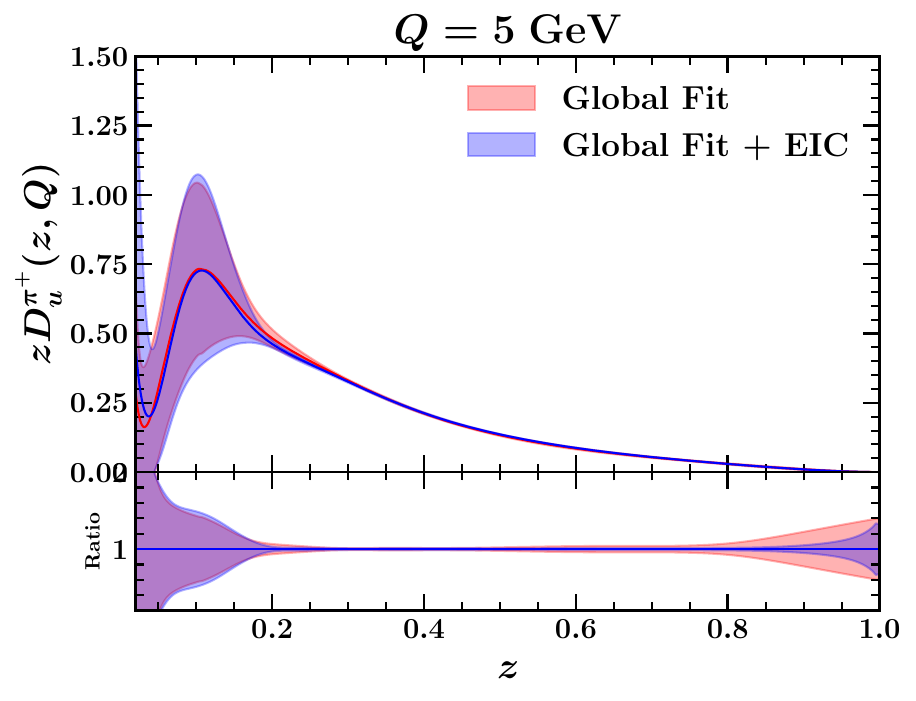}}
\subfloat{\includegraphics[width=0.39\textwidth]{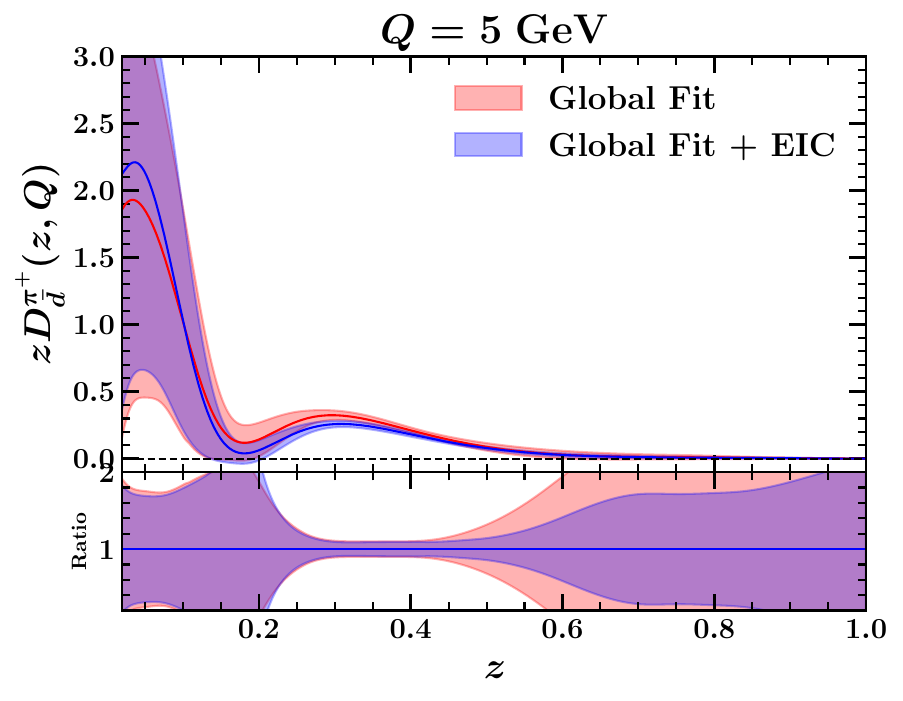}} \\ 
\subfloat{\includegraphics[width=0.39\textwidth]{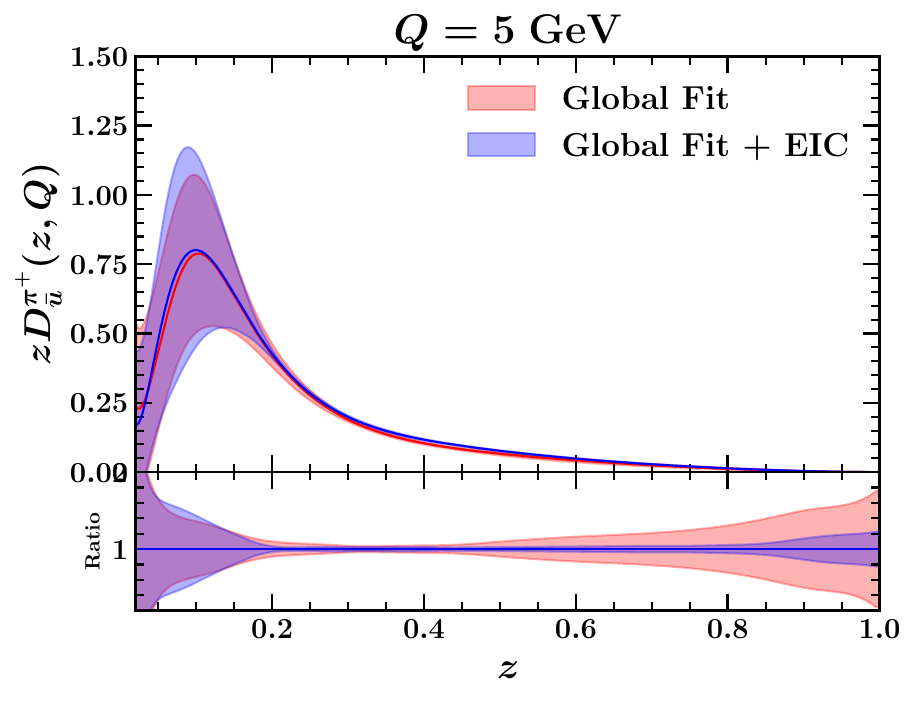}} 
\subfloat{\includegraphics[width=0.39\textwidth]{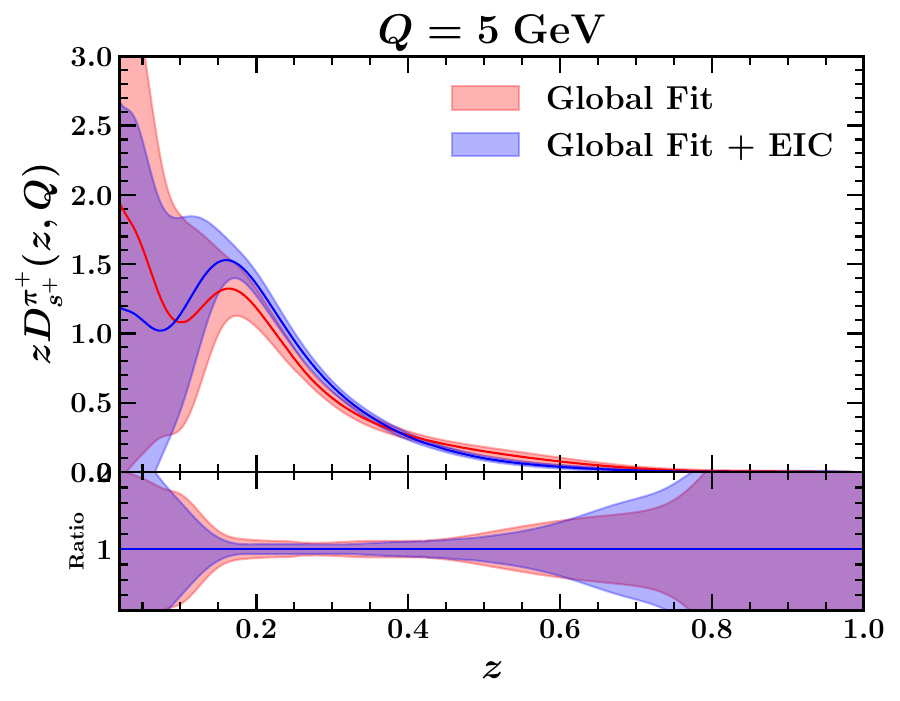}} \\ 
\subfloat{\includegraphics[width=0.39\textwidth]{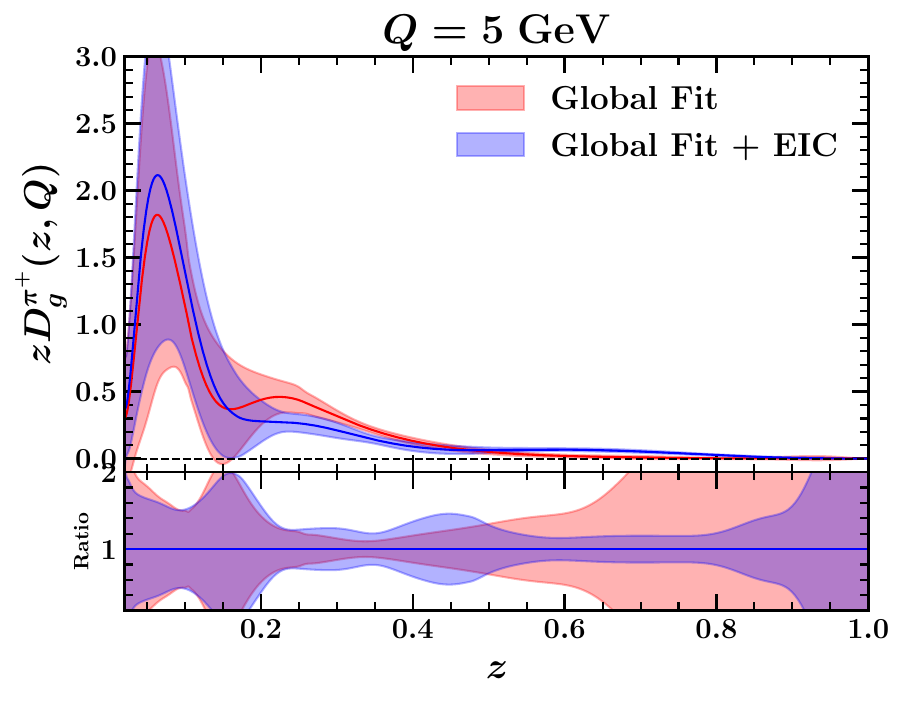}}	
\subfloat{\includegraphics[width=0.39\textwidth]{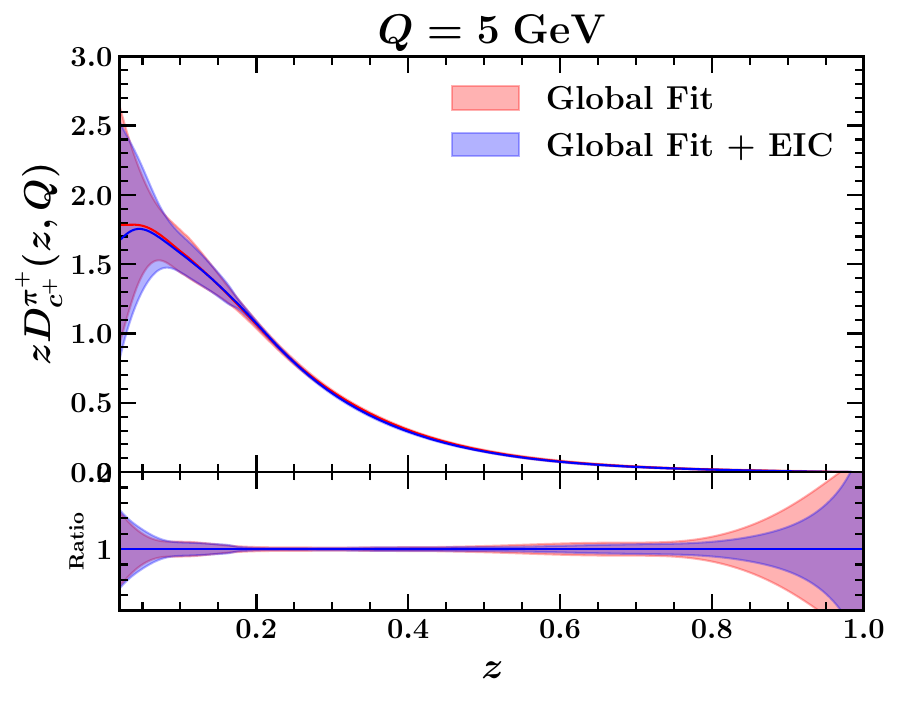}}\\	
\subfloat{\includegraphics[width=0.39\textwidth]{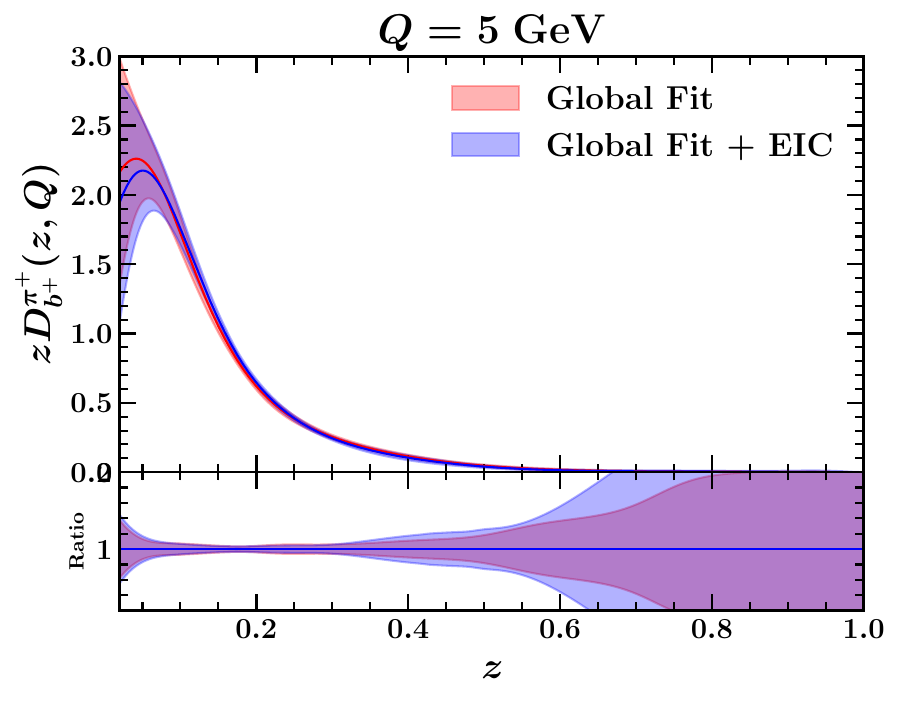}}	
\begin{center}
\caption{\small
Comparison of the $\pi^+$ FFs at NLO accuracy for various parton species at the 
initial scale $Q = 5$~GeV, extracted from a global fit without EIC data (red bands) 
and with EIC pseudo-data included (blue bands). The results are shown FFs in the upper panels for the up quark ($D^u_{\pi^+}$), 
anti-down quark ($D^{\bar{d}}_{\pi^+}$), anti-up quark ($D^{\bar{u}}_{\pi^+}$),  
strange quark ($D^s_{\pi^+}$),  the gluon ($D^g_{\pi^+}$) and heavy-quark FFs. Lower panels represent the ratios of the all sets to the central values of the corresponding sets.}  
\label{fig:FFsNLO}
\end{center}
\end{figure*}

Figure~\ref{fig:FFsNLO} presents a comparison of the extracted pion FFs at NLO accuracy for various parton species 
at the initial scale $Q = 5$~GeV from both `Global Fit' and `Global Fit + EIC'. The upper panel of each plot shows the absolute distributions, while the lower panel displays the ratios relative to the central values of the corresponding fits.
The results are shown FFs for the up quark ($D^u_{\pi^+}$), anti-down quark ($D^{\bar{d}}_{\pi^+}$), anti-up quark ($D^{\bar{u}}_{\pi^+}$), and 
strange quark ($D^s_{\pi^+}$). The figure also displays the gluon ($D^g_{\pi^+}$) and heavy-quark FFs.

From Figure~\ref{fig:FFsNLO}, several key observations can be made. 
We focus only on the reduction of uncertainties in the extracted FFs when EIC pseudo--data are included in the fit. The shaded bands show that these uncertainties become much smaller with the addition of the EIC data, as expected from their projected precision.
 This reduction is particularly evident for the up-quark ($D^u_{\pi^+}$), 
anti-up quark ($D^{\bar{u}}_{\pi^+}$) and anti-down quark ($D^{\bar{d}}_{\pi^+}$) FFs, emphasizing that future EIC measurements substantially 
enhance the precision of favored and unfavored quark FFs, particularly in the currently poorly constrained high-$z$ region.

The gluon FF ($D^g_{\pi^+}$) also shows improvements in its uncertainty band, 
although this reduction is more modest compared to the quark FFs, 
reflecting ongoing challenges in accurately constraining gluon fragmentation. 
Overall, these comparisons illustrate that the inclusion of EIC pseudo-data notably enhances the precision of FFs.
This improvement underscores the critical role future EIC measurements will play in refining global QCD analyses of FFs.

\subsection{Comparison with Other FF Sets}\label{Comparison}
\begin{figure*}[htb]
\vspace{0.50cm}
\centering    
\subfloat{\includegraphics[width=0.4\textwidth]{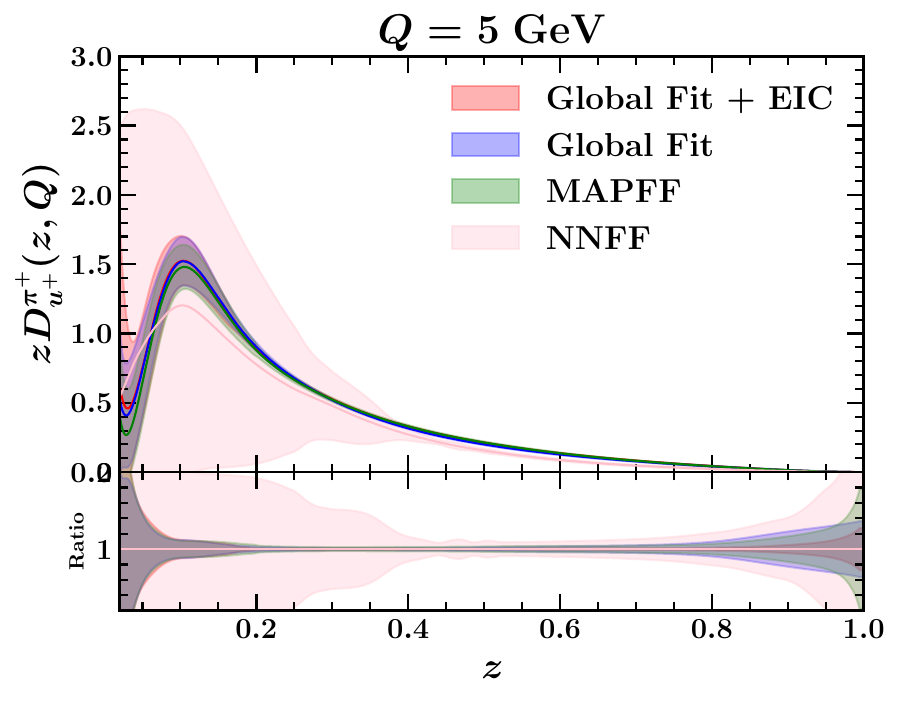}} 	
\subfloat{\includegraphics[width=0.4\textwidth]{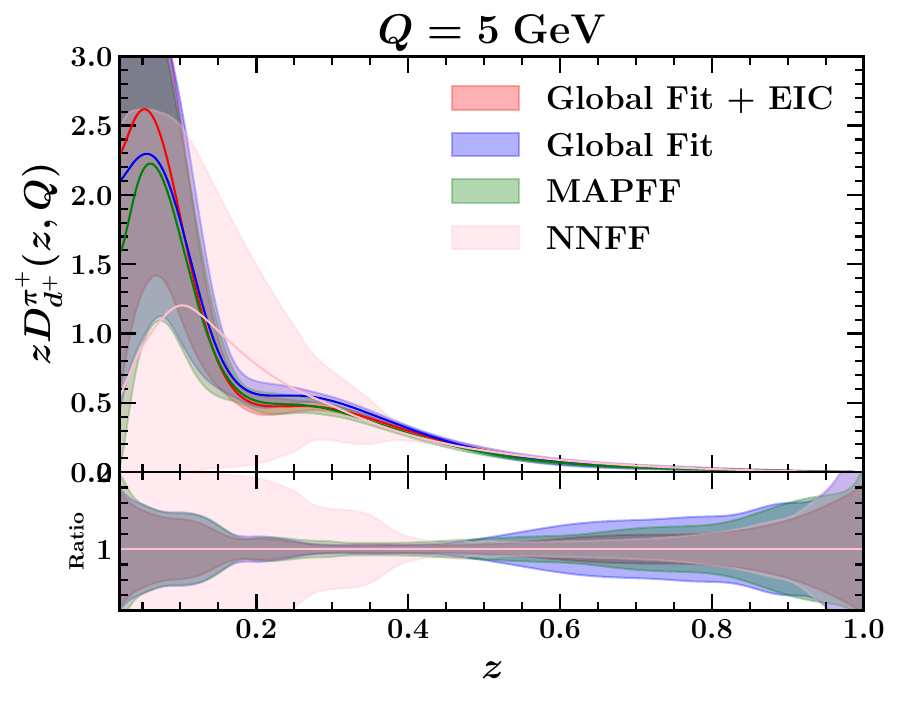}}\\
\subfloat{\includegraphics[width=0.4\textwidth]{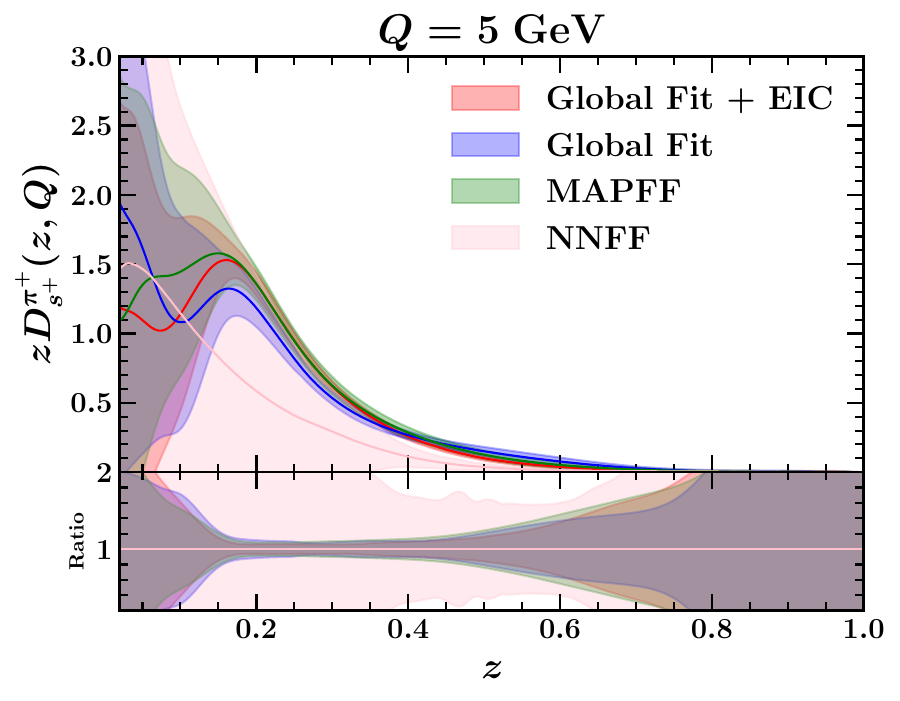}}
\subfloat{\includegraphics[width=0.4\textwidth]{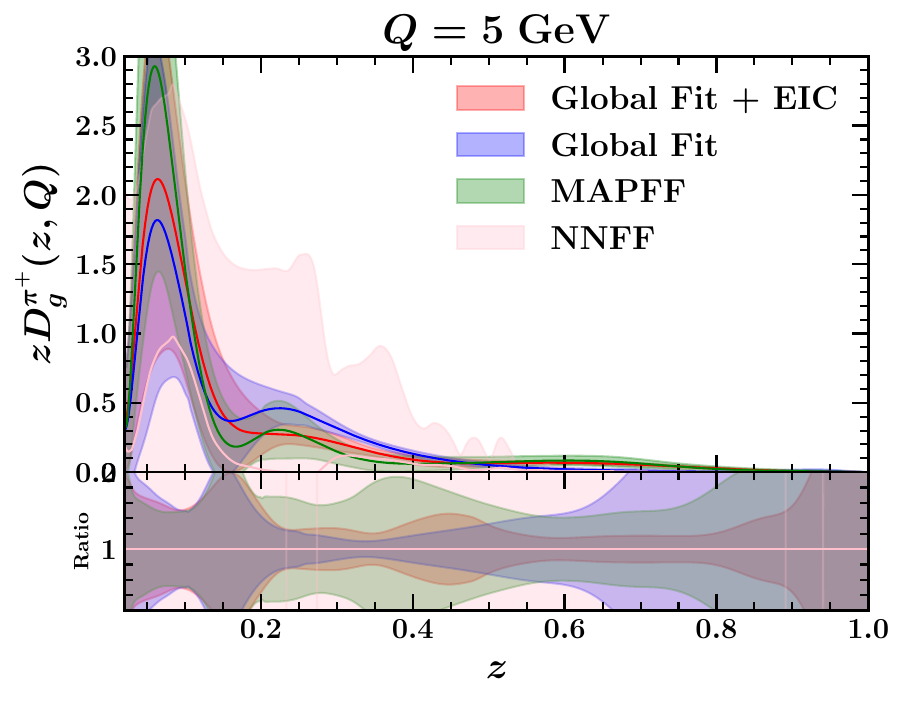}}		
\begin{center}
\caption{\small
Comparison of pion ($\pi^+$) FFs at NLO accuracy for various parton species at the initial scale $Q = 5$~GeV, obtained from our global fit with and without 
including EIC pseudo-data (red and blue bands respectively), the MAPFF extraction~\cite{AbdulKhalek:2022laj} (green bands) and the NNFF1.0 analysis~\cite{Bertone:2017tyb} (pink bands). 
The upper plots show FFs for the 
up-type quark combination ($D^{u^+}_{\pi^+}$) and down-type quark combination ($D^{d^+}_{\pi^+}$), while the 
lower plots display the FFs for the strange quark combination ($D^{s^+}_{\pi^+}$) and gluon ($D^{g}_{\pi^+}$). Lower panels represent the ratios of the all sets to the central values of the corresponding sets.
The shaded bands represent the $1\sigma$ uncertainty intervals, 
computed using the Monte Carlo sampling method. }
\label{fig:compare_FFsNLO}
\end{center}
\end{figure*}

Now we compare our extracted pion FFs with two other existing sets available in the literature. Our global QCD analyses, with and without
incorporating EIC pseudo-data (labeled ``Global Fit + EIC'' and ``Global Fit''), are compared with 
the MAPFF extraction~\cite{AbdulKhalek:2022laj} and results from the NNFF1.0 collaboration~\cite{Bertone:2017tyb}.
Figure~\ref{fig:compare_FFsNLO} presents a detailed comparison among these four sets of pion FFs at NLO accuracy and at 
the initial scale $Q = 5$~GeV. The upper panel of each plot shows the absolute distributions of the set, while the lower panel displays the ratios relative to the central values of the corresponding sets.

Our "Global Fit + EIC" and "Global Fit" analyses in comparison with the MAPFF extraction, utilize similar datasets, 
with the primary differences being the inclusion of EIC pseudo-data 
in our "Global Fit + EIC"  analysis and slight variations in the kinematic cuts applied to the HERMES and COMPASS SIDIS datasets. The kinematic cut on the virtuality Q in MAPFF is $Q_{cut}=2$~GeV to the HERMES and COMPASS SIDIS datasets while it is $Q_{cut}=1.4$~GeV in our analyses for all SIDIS datasets. Moreover, MAPFF has used the NNPDF3.1 \cite{NNPDF:2017mvq} PDF sets as input to the computation of SIDIS cross sections. However we use the NNPDF4.0 \cite{NNPDF:2024dpb} PDF sets.
In contrast, the NNFF1.0 QCD fit exclusively uses SIA datasets. 
The upper plots in Figure~\ref{fig:compare_FFsNLO} show the FFs for the up-type quark combination ($D^{u^+}_{\pi^+}$) and down-type quark combination ($D^{d^+}_{\pi^+}$), 
while the lower plots display the FFs for the strange quark combination ($D^{s^+}_{\pi^+}$) and gluon ($D^{g}_{\pi^+}$).

Comparison of our "Global Fit" results with the MAPFF analysis shows similar behavior in both the central values and the error bands for all types of light partons. However, for the gluon fragmentation function, significant differences are observed between our baseline analysis ("Global Fit") and the MAPFF results across the entire range of $z$, particularly at low $z$ which can be related to difference of kinematic cuts for $Q$ of SIDIS data sets in these two analyses.
Comparing our "Global Fit + EIC" results with those of the MAPFF analysis, we observe largely consistent central 
values across all parton species, owing to similarities in flavor decomposition methods. 
The uncertainty bands, indicated by shaded regions, differ noticeably between these two extractions. 
For the quark FFs $D^{u^+}_{\pi^+}$ and $D^{d^+}_{\pi^+}$, already constrained well by existing data, 
uncertainties are relatively small and similar between both analyses. 
Nevertheless, incorporating EIC pseudo-data leads to additional uncertainty reduction at high-$z$, 
highlighting improved constraints in this kinematic region. 
More substantial improvements are observed in the gluon FF ($D^g_{\pi^+}$), especially at intermediate and high $z$. 
The inclusion of EIC pseudo-data significantly narrows the gluon uncertainty bands, 
underscoring the critical impact of future EIC measurements in precisely constraining gluon fragmentation.

Comparisons between our "Global Fit + EIC" results and the NNFF1.0 extraction reveal substantial differences in both 
central values and uncertainties, especially at medium and lower $z$. 
These discrepancies primarily result from NNFF1.0's exclusive reliance on SIA datasets, 
limiting its ability to constrain FFs effectively. Consequently, NNFF1.0 uncertainties are considerably 
broader compared to our analysis, which benefits from additional constraints provided by SIDIS and projected EIC measurements.

Overall, this comparative analysis emphasizes the significant impact anticipated from incorporating EIC pseudo-data in global FFs extractions. 
It particularly highlights expected enhancements for gluon, u-quark, and anti-u-quark FFs compared to other FFs sets.

\section{Summary and Outlook}\label{summary}

Semi-inclusive deep-inelastic scattering offers a powerful and versatile means of 
probing both the flavor structure of the nucleon and the mechanisms through which partons hadronize into observed final-state hadrons. 
Utilizing the framework of QCD factorization, the SIDIS cross-sections can be accurately described 
in terms of non-perturbative fragmentation functions, thereby providing stringent constraints 
on their extraction from experimental measurements.

In this study, we performed a detailed analysis to quantify the anticipated impact of 
future SIDIS measurements from the Electron-Ion Collider on the determination of pion FFs. 
By conducting a global QCD fit including existing SIA and SIDIS data, complemented by 
simulated pseudo-data expected from the EIC SIDIS, we have explicitly demonstrated how forthcoming EIC 
measurements will enhance the precision of FF extractions.

Our analysis reveals that the inclusion of EIC pseudo-data leads to substantial improvements in 
the precision of FFs, especially in the poorly constrained regions of large momentum fractions ($z$). 
We found that the uncertainty bands, particularly for 
the gluon and up-quark FFs, show significant reductions upon inclusion of EIC pseudo-data.
This highlights the crucial role of future EIC SIDIS measurements in constraining quark and gluon FFs, addressing a 
long-standing challenge in fragmentation phenomenology.

Comparisons with the SIA and SIDIS-based extraction indicate that the EIC pseudo-data significantly 
tighten constraints on gluon FFs, underscoring the potential of the EIC to advance the current 
state-of-the-art in FF precision. 
This comparative analysis clearly emphasizes that SIDIS measurements at the EIC will 
complement existing data by probing a wider kinematic range, thereby improving our knowledge of hadronization mechanisms.

In the near future, we plan to extend this study to Kaon FFs. 
Given the increased sensitivity of Kaon production to strange quark distributions, this analysis will 
enhance our understanding of flavor-dependent fragmentation processes.

Additionally, we will explore the impact of new data from the BESIII experiment on the production of charged pions ($\pi^\pm$) 
in electron-positron annihilation~\cite{BESIII:2025mbc}. We will also examine the recently reported multiplicities for charged pion 
production in SIDIS on proton and deuteron targets from the JLab collaboration~\cite{Bhatt:2024prq}.

\acknowledgments

The authors gratefully acknowledge the helpful discussions provided by Charlotte Van Hulse. 
We also thank the School of Particles and Accelerators at the Institute for Research in Fundamental Sciences
(IPM) for their financial support of this project. 
Hamzeh Khanpour appreciates the financial support from NAWA under grant number BPN/ULM/2023/1/00160, as well as 
from the IDUB program at AGH University.



%

\end{document}